\documentclass[journal]{IEEEtran} 
\IEEEoverridecommandlockouts
\usepackage{cite}
\usepackage{amsmath,amssymb,amsfonts}
\usepackage{algorithmic}
\usepackage{graphicx}
\usepackage{multirow}
\usepackage{textcomp}
\usepackage{hyperref}
\usepackage{xcolor}
\usepackage{comment}
\def\BibTeX{{\rm B\kern-.05em{\sc i\kern-.025em b}\kern-.08em
    T\kern-.1667em\lower.7ex\hbox{E}\kern-.125emX}}

\newcommand{\mysmallarraydecl}{\renewcommand{%
\IEEEeqnarraymathstyle}{\scriptscriptstyle}%
\renewcommand{\IEEEeqnarraytextstyle}{\tiny}%
\renewcommand{\baselinestretch}{1.1}%
\settowidth{\normalbaselineskip}{\scriptsize
\hspace{\baselinestretch\baselineskip}}%
\setlength{\baselineskip}{\normalbaselineskip}%
\setlength{\jot}{0.25\normalbaselineskip}%
\setlength{\arraycolsep}{2pt}}
\begin{document}

\title{The Art NFTs and Their Marketplaces\\
}

\author{
    \IEEEauthorblockN{Lanqing Du\IEEEauthorrefmark{1}, Michelle Kim\IEEEauthorrefmark{2}, Jinwook Lee\IEEEauthorrefmark{1}}
    \\
    \IEEEauthorblockA{\IEEEauthorrefmark{1} \textit{Drexel University, Philadelphia, PA 19104, USA}}\\
    \IEEEauthorblockA{\IEEEauthorrefmark{2} \textit{Horace Mann School, Bronx, NY 10471, USA}}\\
    \IEEEauthorblockA{Email: jl3539@drexel.edu (Corresponding Author: Jinwook Lee)}
}

\maketitle

\begin{abstract}
    Non-Fungible Tokens (NFTs) are crypto assets with a unique digital identifier for ownership, powered by blockchain technology. Technically speaking, anything digital could be minted and sold as an NFT, which provides proof of ownership and authenticity of a digital file. For this reason, it helps us distinguish between the originals and their copies, making it possible to trade them. This paper focuses on art NFTs that change how artists can sell their products. The art NFTs also change how the art trade market works since NFT technology cuts out the middleman. Recently, the utility of NFTs has become an essential issue in the NFT ecosystem, which refers to the owners' usefulness, profitability, and benefits. Using recent major art NFT marketplace datasets, we summarize and interpret the current market trends and patterns in a way that brings insight into the future art market. Numerical examples are presented.
\end{abstract}


\section{Introduction}
\subsection{NFTs and the Art Industry}
In April, 2022, Sotheby’s sold a small receipt paper of the 1959 project called “Zone of Empty Space” by Yves Klein, the French conceptual artist, for \$1.2 million. It was a part of the ledger where Klein recorded all sales and resales of the 1959 artwork (\cite{b11}). More than a half-century later, thanks to blockchain technology and NFTs, this ledger-keeping has become an essential part of the art industry. (\cite{b1, b2}) 

The phrase “NFT art,” we believe, is not the most accurate phrase. NFT itself is not the art, it is simply a technology that increases the utility of the art, by functioning as a proof and traceability of the ownership. (\cite{b3, b4, b5}) Thus, throughout the rest of the paper, we will be using the phrase “art NFT” as opposed to “NFT art.”
There are three types of art NFT: digital art (stand-alone), PFP (generative art), and Phygital art (linking physical art with the NFT). (\cite{b6, b7, b8}) The very brief history of NFTs begins with digital art and its pioneer Kevin McCoy. (\cite{b9, b10}) In 2014 McCoy and Anil Dash created the first stand-alone NFT, Quantum.(\cite{b12}) Prior to Quantum, digital artworks were “fungible,” meaning that there were multiples of the same artwork. Following McCoy, artists like Mike Winkelmann, better known as Beeple, and companies like Larva Labs, creator of CryptoPunks, took center stage on the NFT market.

The launch of CryptoPunks marked the creation of a new category of art NFT: PFP (Profile Pic) created using a technology called generative art. PFP art including CryptoPunks, Bored Ape Yacht Club (BAYC), Doodles, and recently Clone X doubled as a form of art and a status symbol on various social media platforms. In the last couple of years, BAYC became widely popular within the NFT community and beyond, with the most expensive piece, \#8817, selling for \$3.4 million in 2021. (\cite{b13}) Yuga Labs, the creator of BAYC, have fully experimented with and implemented business strategy models like token-gating.
Token-gating is a way of adding value to an NFT by granting the holder exclusive access to content, community, events, and physical products, in addition to the digital token. (\cite{b14}) BAYC was also the first art that granted full commercial rights to the Intellectual Property (IP) to its holders, who were now able to commercialize their Bored Apes. Unsurprisingly, popularity factors for BAYC include commercial rights and exclusive access to spin-off collections like Bored Ape Kennel Club (BAKC) and Mutant Ape Yacht Club (MAYC) both of which have high resale value as well access to off-line events including the annual Ape Fest.

If commerce platforms like Nifty Gateway and Superway became well-known for their curation of digital art, marketplaces like OpenSea became highly successful from their listing of PFP art, such as the BAYC. While digital art centers around individual artists, like McCoy and Beeple, PFP centers around companies, Yuga Labs and Larva Labs being the most notable, where they curate communities, introduce roadmaps and coordinate on and offline events and launches.

There are not yet any notable cases of the third type of art NFT, ``Phygital" art (Physical and Digital art). Yet, by linking NFT (digital proof of authenticity) with physical art, it will revolutionize both the NFT and the traditional art market. The prime challenge is finding an optimal method of linking the two; Quick Response (QR) codes, Radio Frequency Identification (RFID), and Near Field Communication (NFC) are a few ways.

Art NFT is changing the landscape of the art market and its players including, the artist, buyer, and platforms (galleries, online commerce platform). While in the traditional art market, galleries and auctions and their agents functioned as an intermediary between the artist, the artwork, and the customers, the emergence of NFTs has bridged the gap between the three. Now the artist or the creator can list and mint their artwork directly on an online platform (with little to no commission fee) and oftentimes connect directly with their buyers. Though there are still technical, ethical, and sometimes legal issues associated with art NFT, it will shift the dynamic between the artist and the buyer, the role of the intermediaries (auctions, galleries, online platforms), and most notably the trends and value surrounding art.

\begin{figure*}[!t]
  \includegraphics[width=\textwidth]{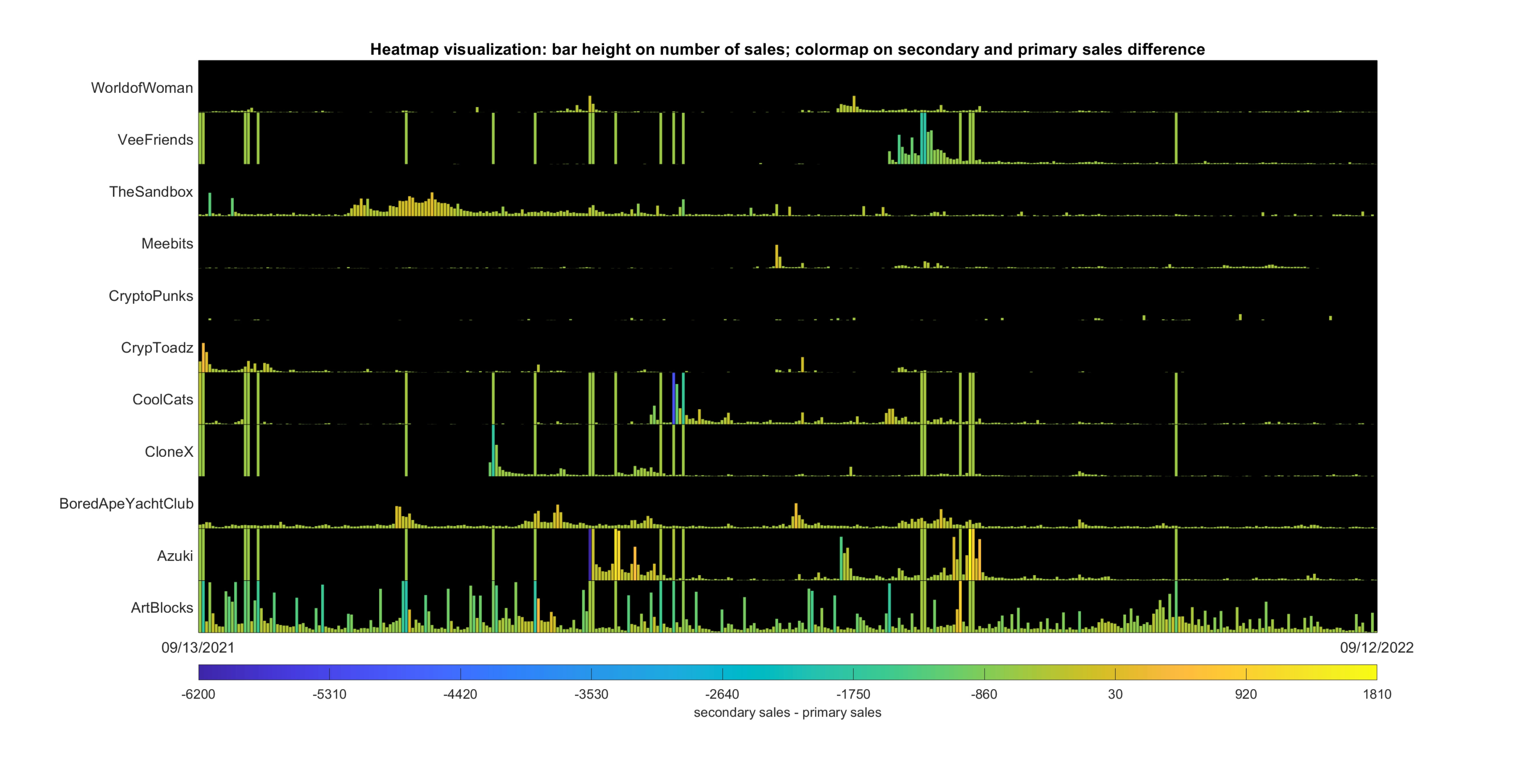}
  \caption{Heatmap visualization: bar height on the number of sales; colormap on secondary and primary sales difference.}
  \label{saleNumSecFirst}
\end{figure*}

\subsection{Data Collection and Our Key Findings}\label{dataDescription}

\begin{figure*}[!t]
  \includegraphics[width=\textwidth]{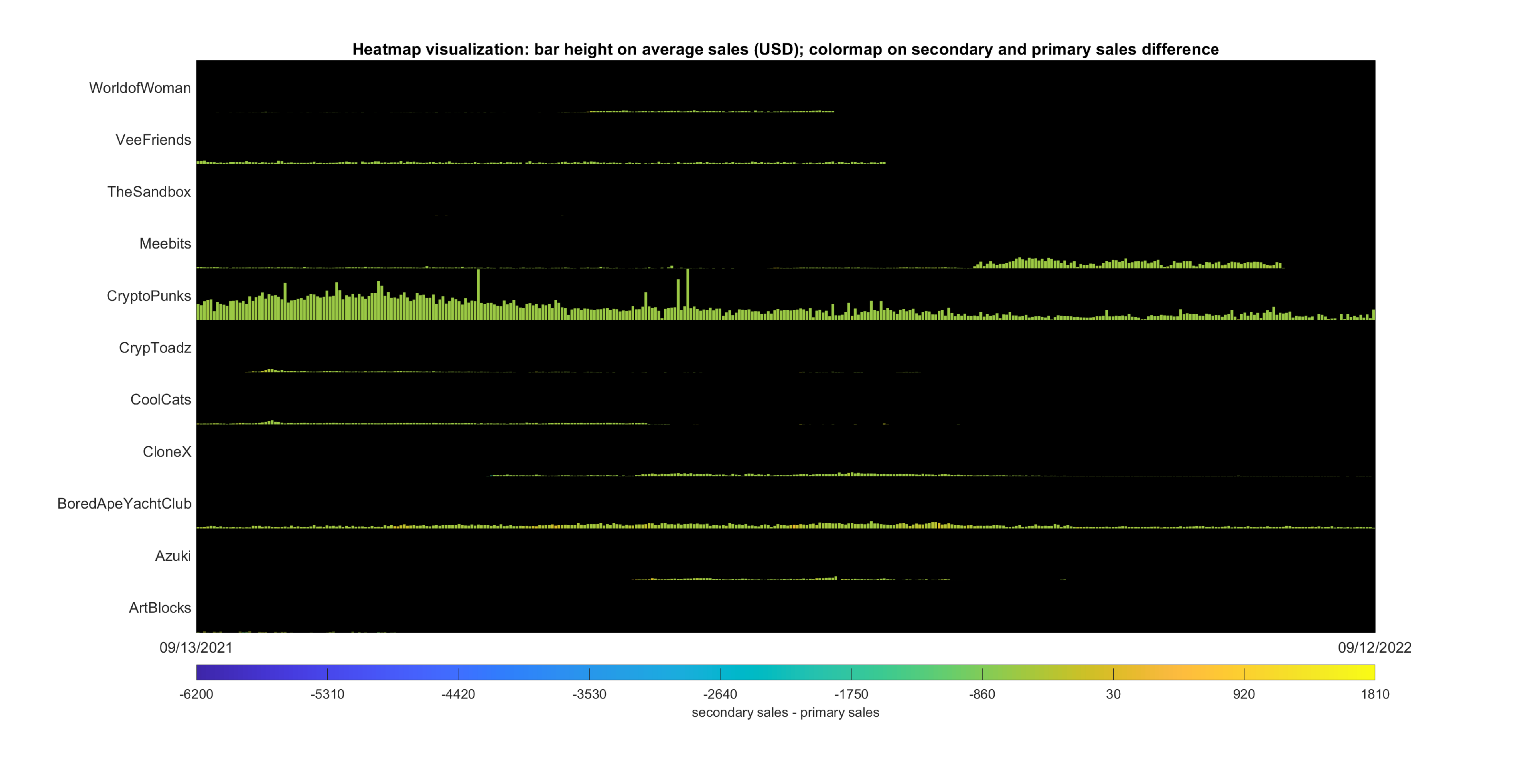}
  \caption{Heatmap visualization: bar height on average sales (in USD); colormap on secondary and primary sales difference.}
  \label{SalesSecFirst}
\end{figure*}

\begin{figure*}[!t]
  \includegraphics[width=\textwidth]{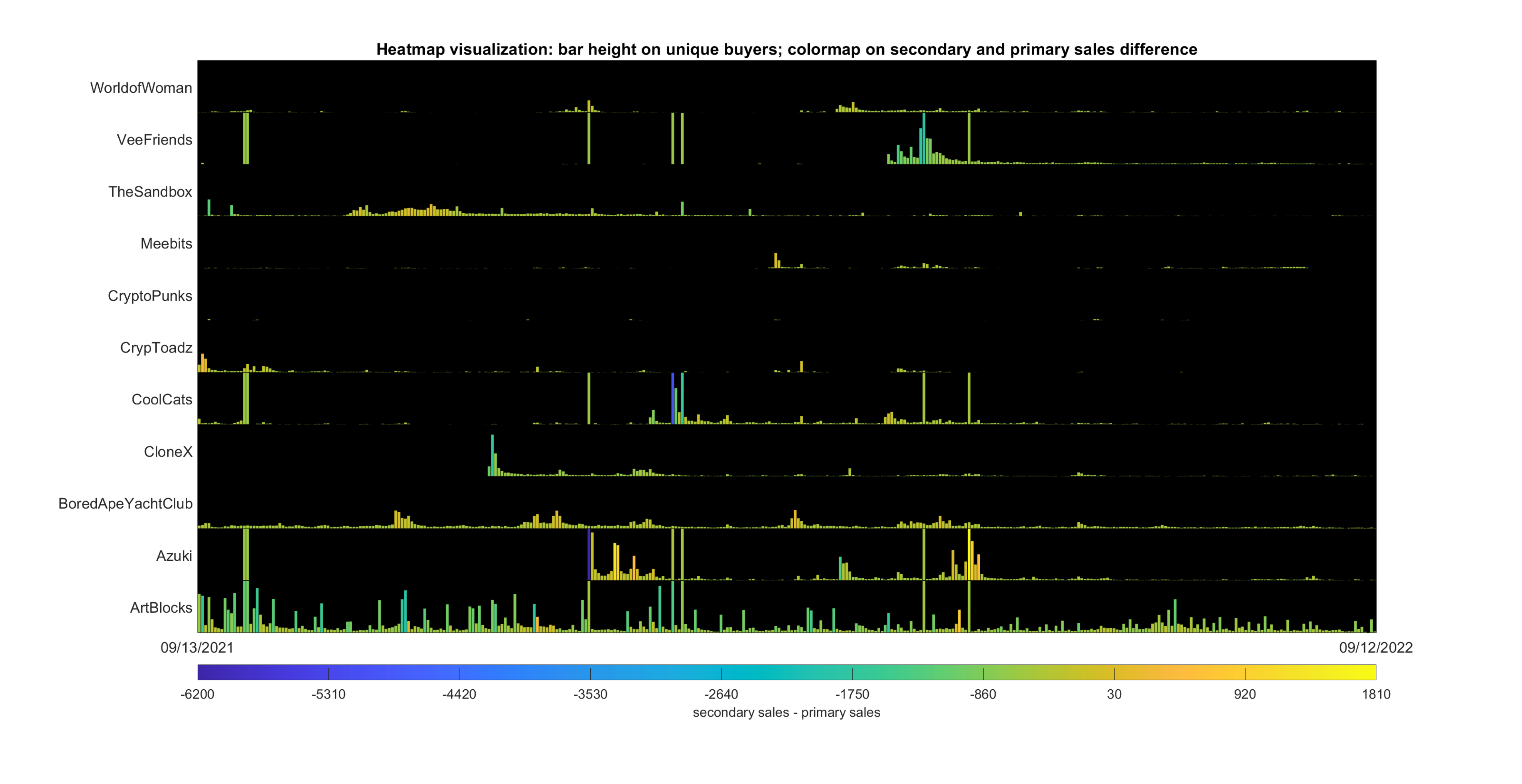}
  \caption{Heatmap visualization: bar height on the number of unique buyers; colormap on secondary and primary sales difference.}
  \label{buyerSecFirst}
\end{figure*}

Our data collection is as follows: 11 marketplaces datasets from NonFungible.com (\cite{b15}): ArtBlocks, Azuki, BoredApeYachtClub, CloneX, CoolCats, CrypToadz, CryptoPunks, Meebits, TheSandbox, VeeFriends, and WorldofWoman. This is daily NFT marketplace datasets with one-year multivariate time series datasets from 09/13/2021 to 09/12/2022. 
Each dataset contains the following ten features: (1) the number of sales (transactions), (2) total sales (USD), (3) average sales (USD), (4) the number of active market wallets, (5) primary sales, (6) secondary sales, (7) primary sales (USD), (8) secondary sales (USD), (9) unique buyers, and (10) unique sellers. Note that null values, $<0.05\%$ of the entire dataset, are deleted in the given dataset.
\begin{itemize}
    \item As shown in Fig. \ref{saleNumSecFirst}, \ref{SalesSecFirst}, and \ref{buyerSecFirst}, the majority of the number of sales (i.e., transactions) comes from secondary sales (approx. 90.16\%), which means that the primary sales are less frequent than the secondary sales.
    \item It appears that primary sales are more frequently observed in ArtBlocks (approx. 30.33\%). While in some specific time periods, primary sales are more often than secondary ones for Coolcats and VeeFriends.
    \item It also appears that there are seasonal patterns as regards buyer activities. As shown in Fig. \ref{saleNumSecFirst} and \ref{buyerSecFirst}, we observed some seasonal patterns for the number of sales and the number of unique buyers.
    \item Speaking of similarity in the trends and patterns for the selected marketplaces, CryptoPunks appears to have fewer but more significant transactions (as depicted in Fig 1 and 4). In comparison, ArtBlocks has more transactions with smaller trading sizes.
\end{itemize}
In Table \ref{tab1}, such art NFT marketplaces are selected based on “All-time NFT Collection Rankings by Sales Volume” from Cryptoslam (\cite{b16}) subject to the data available in the NonFungible.com database.

\begin{table}[htbp]
    \caption{NFT Collection Rankings by Sales Volume}
    \begin{center}
        \begin{tabular}{cccc}
            \hline
Collection & Sales & Buyers & Txns\\
\hline
Axie Infinity & \$4,087M & 1,785,424 & 1,785,424\\
Bored Ape Yaght Club & \$2,416M & 11,977 & 32,465\\
CryptoPunks & \$2,371M & 6,033 & 22,157\\
Mutant Ape Yacht Club & \$1,731M & 23,513 & 51,111\\
Art Blocks & \$1,305M & 33,113 & 181,667\\
Otherdeed & \$1,047M & 25,854 & 58,319 \\
NBA TopShot & \$1,030M & 445,999 & 21,695,225\\ 
            \hline
        \end{tabular}
        \label{tab1}
    \end{center}
\end{table}

\section{Machine learning on the NFT marketplace datasets}

\subsection{Unsupervised Learning: PCA}
The principal components of the data matrix are its singular vectors. Using the SVD (Singular Value Decomposition), Principal Component Analysis (PCA) finds the largest singular values to extract the most important information from the data (with the largest variance) by solving perpendicular least squares (i.e., orthogonal regression). PCA, in machine learning, is unsupervised learning. PCA is typically carried out to generate the principal components (PCs), i.e., singular vectors, to summarize a large dataset with correlated variables (see, e.g., \cite{b16}). 
Below, (1) describes the correlation between 10 features for the art NFT marketplace BAYC, with F1 standing for the first feature (number of sales) and F10 for the last feature (unique sellers) as introduced in section \ref{dataDescription}.

\begin{equation}
\label{eq1}
\left(\begin{IEEEeqnarraybox*}[\mysmallarraydecl][c]{,c/c/c/c/c/c/c/c/c/c/c,}
& F1 & F2 & F3 & F4 & F5 & F6 & F7 & F8 & F9 &F10\\  
F1 & 1.00 & 0.91 & 0.21 & 0.99 & 0.81 & 0.99 & 0.83 & 0.90 & 0.99 & 0.99 \\ 
F2 & 0.91 & 1.00 & 0.51 & 0.88 & 0.61 & 0.92 & 0.78 & 0.99 & 0.89 & 0.89 \\ 
F3 & 0.21 & 0.51 & 1.00 & 0.21 & 0.03 & 0.22 & 0.29 & 0.51 & 0.21 & 0.21 \\  
F4 & 0.99 & 0.88 & 0.21 & 1.00 & 0.84 & 0.98 & 0.84 & 0.87 & 0.99 & 0.99 \\   
F5 & 0.81 & 0.61 & 0.03 & 0.84 & 1.00 & 0.77 & 0.84 & 0.59 & 0.83 & 0.83 \\   
F6 & 0.99 & 0.92 & 0.22 & 0.98 & 0.77 & 1.00 & 0.81 & 0.91 & 0.99 & 0.99 \\   
F7 & 0.83 & 0.78 & 0.29 & 0.84 & 0.84 & 0.81 & 1.00 & 0.75 & 0.84 & 0.84 \\   
F8 & 0.90 & 0.99 & 0.51 & 0.87 & 0.59 & 0.91 & 0.75 & 1.00 & 0.88 & 0.89 \\  
F9 & 0.99 & 0.87 & 0.21 & 0.99 & 0.83 & 0.98 & 0.84 & 0.88 & 1.00 & 0.99 \\  
F10 & 0.99 & 0.89 & 0.22 & 0.99 & 0.83 & 0.98 & 0.84 & 0.89 & 0.99 & 1.00
\end{IEEEeqnarraybox*}\right)
\end{equation}

The correlation coefficients matrix of (\ref{eq1}) shows that the BAYC dataset is highly and positively correlated. Most of our selected NFT marketplaces have similar correlation structures, while CryptoPunks in (\ref{eq2}) only is different from others: (i) less strong correlation as in (\ref{eq2}); (ii) negative correlation between average sales and both primary and secondary sales. (also refer to Fig.\ref{saleNumSecFirst} and \ref{buyerSecFirst}).

\begin{equation}
\label{eq2}
\left(\begin{IEEEeqnarraybox*}[\mysmallarraydecl][c]{,c/c/c/c/c/c/c/c/c/c/c,}
& F1 & F2 & F3 & F4 & F5 & F6 & F7 & F8 & F9 &F10\\ 
F1 & 1.00  & 0.44  & -0.15 & 0.47  & 0.42  & 0.97  & 0.33  & 0.43  & 0.47  & 0.49 \\
F2 &    0.44  & 1.00  & 0.56  & 0.70  & 0.17  & 0.44  & 0.50  & 0.99  & 0.70  & 0.70 \\
F3 &    -0.15 & 0.56  & 1.00  & 0.05  & -0.14 & -0.13 & 0.14  & 0.57  & 0.05  & 0.04 \\
F4 &    0.47  & 0.70  & 0.05  & 1.00  & 0.12  & 0.48  & 0.35  & 0.70  & 0.99  & 0.99 \\
F5 &    0.42  & 0.17  & -0.14 & 0.12  & 1.00  & 0.20  & 0.68  & 0.09  & 0.13  & 0.12 \\
F6 &    0.97  & 0.44  & -0.13 & 0.48  & 0.20  & 1.00  & 0.18  & 0.44  & 0.48  & 0.49 \\
F7 &    0.33  & 0.50  & 0.14  & 0.35  & 0.68  & 0.18  & 1.00  & 0.40  & 0.35  & 0.33 \\
F8 &    0.43  & 0.99  & 0.57  & 0.70  & 0.09  & 0.44  & 0.40  & 1.00  & 0.70  & 0.70 \\
F9 &    0.47  & 0.70  & 0.05  & 0.99  & 0.13  & 0.48  & 0.35  & 0.70  & 1.00  & 0.97 \\
F10 &    0.49  & 0.70  & 0.04  & 0.99  & 0.12  & 0.49  & 0.33  & 0.70  & 0.97  & 1.00
\end{IEEEeqnarraybox*}\right)
\end{equation}

Note that, in the given datasets, each feature has its own scale with varying magnitudes, therefore we normalize the datasets before passing them to PCA. {\it Python 3} (version 3.7.14) is used as the programming language and {\it sklearn.decomposition.} PCA function from Scikit-learn package (version 1.0.2) (in \cite{b17}) is applied.

\subsection{Numerical Results}

\begin{figure}[!t]
  \includegraphics[scale = 0.78]{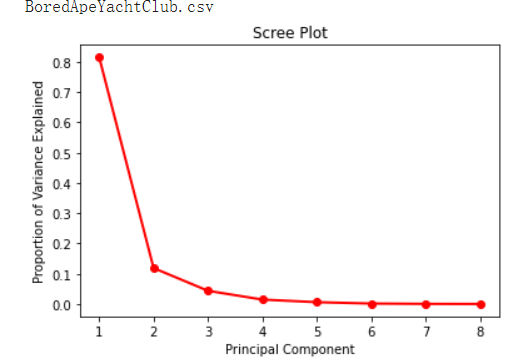}
  \caption{Scree plot for BoredApeYachtClub.}
  \label{screeBAYC}
\end{figure}

\begin{figure}[!t]
  \includegraphics[scale = 0.8]{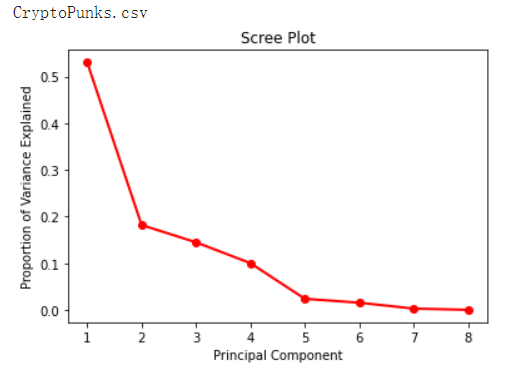}
  \caption{Scree plot for CryptoPunks.}
  \label{screeCrypto}
\end{figure}

Researchers can utilize the scree plot, Kaiser's rule, and/or Horn's procedure (\cite{b16}) with a specified threshold of explained variance to decide how many primary components should be preserved. Fig \ref{screeBAYC} and \ref{screeCrypto} are the scree plots for BAYC and CryptoPunks, respectively. From the Figures, we observe that there are two PCs for BAYC and four PCs for CryptoPunks.
In our numerical analysis, we keep a given number of principal components via scree plot with the following two constraints:
\begin{itemize}
    \item [(i)] Cumulative proportion of variance explained is greater than 90\%
   \item  [(ii)] Eigenvalue of each principal component is $> 1$.
\end{itemize}

Using the above criteria, we retained the following number of principal components as shown in Table \ref{tab2}.

\begin{table}[htbp]
    \caption{NFT Collection Rankings by Sales Volume}
    \begin{center}
        \begin{tabular}{cc}
            \hline
\# of Principal  & \multirow{2}{*}{Art NFT Marketplaces}\\
Components & \\
\hline
\multirow{2}{*}{2} & ArtBlocks, BoredApeYachtClub, CloneX, CoolCats, \\
 & Meebits, TheSandbox, VeeFriends, WorldofWoman\\
3 & Azuki, CrypToadz\\
4 & CryptoPunks\\
            \hline
        \end{tabular}
        \label{tab2}
    \end{center}
\end{table}

Table \ref{eq3} lists the original PCA results for ArtBlocks. In Table (\ref{eq3}), two linear combinations of the given ten features (F1, F2, ..., F10) are presented to interpret ArtBlock’s two principal components.

\begin{table}[!t]
\renewcommand{\arraystretch}{1}
\caption{ArtBlocks Principle Components}
\label{eq3}
\centering
\tiny
\begin{tabular}{ccccccccccc}
\hline
\bfseries & F1  &  F2  &  F3  &  F4  &  F5  &  F6  &  F7  &  F8  &  F9  &  F10\\
\hline
 PC1 &  0.35   &  0.32    &  0.09    &  0.37   &  0.31   &  0.33   &  0.29    &  0.31    &  0.36   &  0.34   \\
 PC2 &  -0.25  &  0.36    &  0.67    &  -0.15  &  -0.29  &  -0.11  &  0.22    &  0.40    &  -0.17  &  -0.09 \\
\hline
\end{tabular}
\end{table}

Table \ref{eq4} - \ref{eq13} list the original PCA results for Azuki, BAYC, CloneX, CoolCats, CrypToadz, CryptoPunks, Meebits, TheSandbox, VeeFriends, and WorldofWoman. 

\begin{table}[!t]
\renewcommand{\arraystretch}{1}
\caption{Azuki Principle Components}
\label{eq4}
\centering
\tiny
\begin{tabular}{ccccccccccc}
\hline
\bfseries & F1  &  F2  &  F3  &  F4  &  F5  &  F6  &  F7  &  F8  &  F9  &  F10\\
\hline
 PC1   & 0.35  & 0.34  & 0.01  & 0.37  & 0.27  & 0.35  & 0.29  & 0.31  & 0.37  & 0.34  \\
 PC2   & 0.28  & -0.25 & -0.36 & -0.06 & 0.52  & -0.16 & 0.40  & -0.43 & 0.10  & -0.27 \\
 PC3   & 0.06  & 0.12  & 0.85  & -0.13 & 0.24  & -0.23 & 0.29  & 0.04  & -0.06 & -0.23 \\
\hline
\end{tabular}
\end{table}

\begin{table}[!t]
\renewcommand{\arraystretch}{1}
\caption{BAYC Principle Components}
\label{eq5}
\centering
\tiny
\begin{tabular}{ccccccccccc}
\hline
\bfseries & F1  &  F2  &  F3  &  F4  &  F5  &  F6  &  F7  &  F8  &  F9  &  F10\\
\hline
& F1 & F2 & F3 & F4 & F5 & F6 & F7 & F8 & F9 & F10 \\PC1  & 0.35  & 0.33  & 0.11  & 0.35  & 0.29  & 0.34  & 0.31  & 0.32  & 0.34  & 0.35 \\PC2  & 0.09  & -0.26 & -0.83 & 0.11  & 0.34  & 0.06  & 0.05  & -0.28 & 0.10  & 0.09 \\
\hline
\end{tabular}
\end{table}

\begin{table}[!t]
\renewcommand{\arraystretch}{1}
\caption{CloneX Principle Components}
\label{eq6}
\centering
\tiny
\begin{tabular}{ccccccccccc}
\hline
\bfseries & F1  &  F2  &  F3  &  F4  &  F5  &  F6  &  F7  &  F8  &  F9  &  F10\\
\hline
PC1    &  0.34    &  0.34    &  0.06    &  0.35    &  0.32    &  0.33    &  0.34    &  0.28    &  0.34    &  0.35     \\PC2    &  -0.14  &  0.14    &  0.84    &  -0.07  &  -0.22  &  0.04    &  -0.09  &  0.43    &  -0.11  &  -0.07 \\
\hline
\end{tabular}
\end{table}


\begin{table}[!t]
\renewcommand{\arraystretch}{1}
\caption{CoolCats Principle Components}
\label{eq7}
\centering
\tiny
\begin{tabular}{ccccccccccc}
\hline
\bfseries & F1  &  F2  &  F3  &  F4  &  F5  &  F6  &  F7  &  F8  &  F9  &  F10\\
\hline
PC1  & 0.35  & 0.32  & -0.02 & 0.36  & 0.33  & 0.34  & 0.33  & 0.26  & 0.35  & 0.35  \\PC2  & -0.13 & 0.35  & 0.73  & -0.10 & -0.17 & 0.01  & -0.14 & 0.51  & -0.13 & -0.01 \\
\hline
\end{tabular}
\end{table}

\begin{table}[!t]
\renewcommand{\arraystretch}{1}
\caption{CrypToadz Principle Components}
\label{eq8}
\centering
\tiny
\begin{tabular}{ccccccccccc}
\hline
\bfseries & F1  &  F2  &  F3  &  F4  &  F5  &  F6  &  F7  &  F8  &  F9  &  F10\\
\hline
PC1  & 0.37  & 0.31  & 0.15  & 0.37  & 0.26  & 0.37  & 0.20  & 0.31  & 0.37  & 0.37  \\PC2  & -0.19 & 0.39  & 0.68  & -0.17 & -0.20 & -0.19 & 0.17  & 0.39  & -0.17 & -0.17 \\ PC3 & -0.10 & -0.15 & 0.02  & -0.11 & 0.57  & -0.10 & 0.75  & -0.17 & -0.11 & -0.12 \\
\hline
\end{tabular}
\end{table}

\begin{table}[!t]
\renewcommand{\arraystretch}{1}
\caption{CryptoPunks Principle Components}
\label{eq9}
\centering
\tiny
\begin{tabular}{ccccccccccc}
\hline
\bfseries & F1  &  F2  &  F3  &  F4  &  F5  &  F6  &  F7  &  F8  &  F9  &  F10\\
\hline
PC1  & 0.30  & 0.38  & 0.10  & 0.39  & 0.13  & 0.29  & 0.23  & 0.37  & 0.39  & 0.39  \\PC2  & -0.43 & 0.27  & 0.58  & 0.04  & -0.41 & -0.36 & -0.13 & 0.30  & 0.04  & 0.03 \\ PC3 & -0.04 & 0.16  & 0.32  & -0.23 & 0.57  & -0.18 & 0.59  & 0.09  & -0.22 & -0.24 \\ PC4 & 0.44  & 0.14  & 0.35  & -0.31 & -0.18 & 0.52  & -0.25 & 0.18  & -0.31 & -0.29 \\
\hline
\end{tabular}
\end{table}

\begin{table}[!t]
\renewcommand{\arraystretch}{1}
\caption{Meebits Principle Components}
\label{eq10}
\centering
\tiny
\begin{tabular}{ccccccccccc}
\hline
\bfseries & F1  &  F2  &  F3  &  F4  &  F5  &  F6  &  F7  &  F8  &  F9  &  F10\\
\hline
PC1  & 0.39  & 0.19  & 0.03  & 0.37  & 0.33  & 0.38  & 0.33  & 0.18  & 0.38  & 0.37   \\PC2  & -0.01 & -0.53 & -0.58 & 0.13  & 0.15  & -0.04 & 0.17  & -0.54 & 0.11  & 0.11\\
\hline
\end{tabular}
\end{table}

\begin{table}[!t]
\renewcommand{\arraystretch}{1}
\caption{TheSandbox Principle Components}
\label{eq11}
\centering
\tiny
\begin{tabular}{ccccccccccc}
\hline
\bfseries & F1  &  F2  &  F3  &  F4  &  F5  &  F6  &  F7  &  F8  &  F9  &  F10\\
\hline
PC1  & 0.36  & 0.36  & 0.22  & 0.37  & 0.16  & 0.34  & 0.20  & 0.34  & 0.36  & 0.36  \\PC2  & 0.10  & -0.14 & -0.28 & 0.05  & 0.65  & -0.19 & 0.53  & -0.26 & 0.20  & -0.19\\
\hline
\end{tabular}
\end{table}

\begin{table}[!t]
\renewcommand{\arraystretch}{1}
\caption{VeeFriends Principle Components}
\label{eq12}
\centering
\tiny
\begin{tabular}{ccccccccccc}
\hline
\bfseries & F1  &  F2  &  F3  &  F4  &  F5  &  F6  &  F7  &  F8  &  F9  &  F10\\
\hline
PC1  & 0.35  & 0.33  & -0.08 & 0.35  & 0.35  & 0.34  & 0.34  & 0.22  & 0.35  & 0.35   \\PC2  & -0.06 & 0.27  & 0.76  & -0.07 & -0.04 & -0.11 & -0.01 & 0.56  & -0.07 & -0.07\\
\hline
\end{tabular}
\end{table}

\begin{table}[!t]
\renewcommand{\arraystretch}{1}
\caption{WorldofWoman Principle Components}
\label{eq13}
\centering
\tiny
\begin{tabular}{ccccccccccc}
\hline
\bfseries & F1  &  F2  &  F3  &  F4  &  F5  &  F6  &  F7  &  F8  &  F9  &  F10\\
\hline
PC1   & 0.35  & 0.31  & 0.00  & 0.35  & 0.31  & 0.35  & 0.31  & 0.30  & 0.35  & 0.35 \\PC2   & -0.13 & 0.38  & 0.70  & -0.14 & -0.29 & -0.10 & 0.24  & 0.39  & -0.13 & -0.13\\
\hline
\end{tabular}
\end{table}

\begin{figure}[!t]
\centering
  \includegraphics[scale = 0.3]{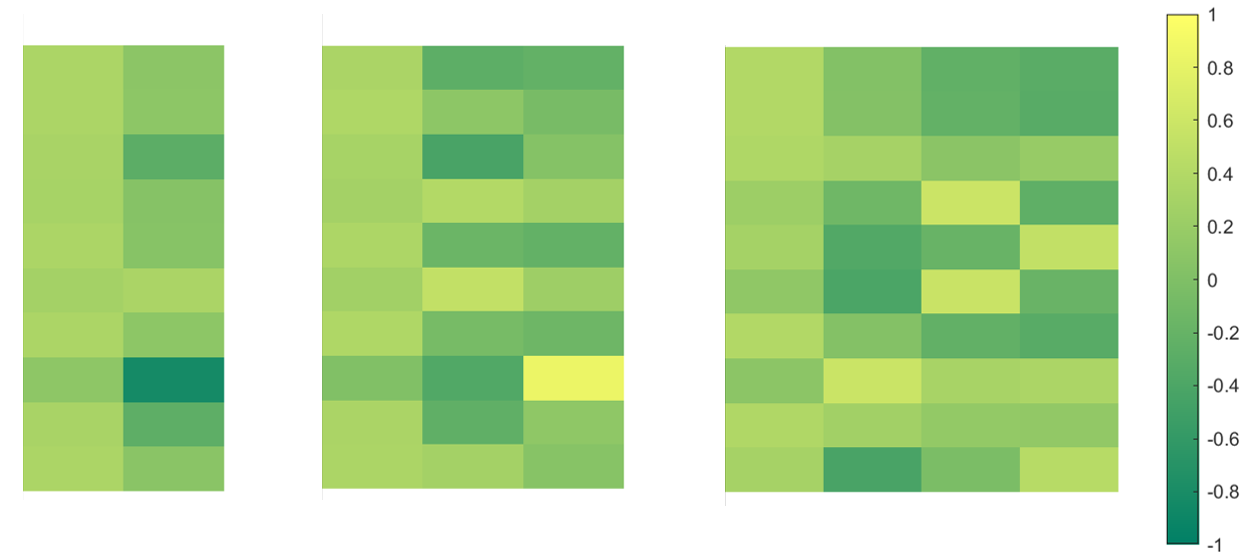}
  \caption{Principal component (PC) coefficient heatmap for NFT with two, three, and four PCs. From left to right: BAYC, Azuki, CryptoPunks.}
  \label{123compare}
\end{figure}

In Fig. \ref{123compare}, for the NFTs with different numbers of PCs, the linear combinations for each PC are different.

\begin{figure}[!t]
\centering
  \includegraphics[scale = 0.4]{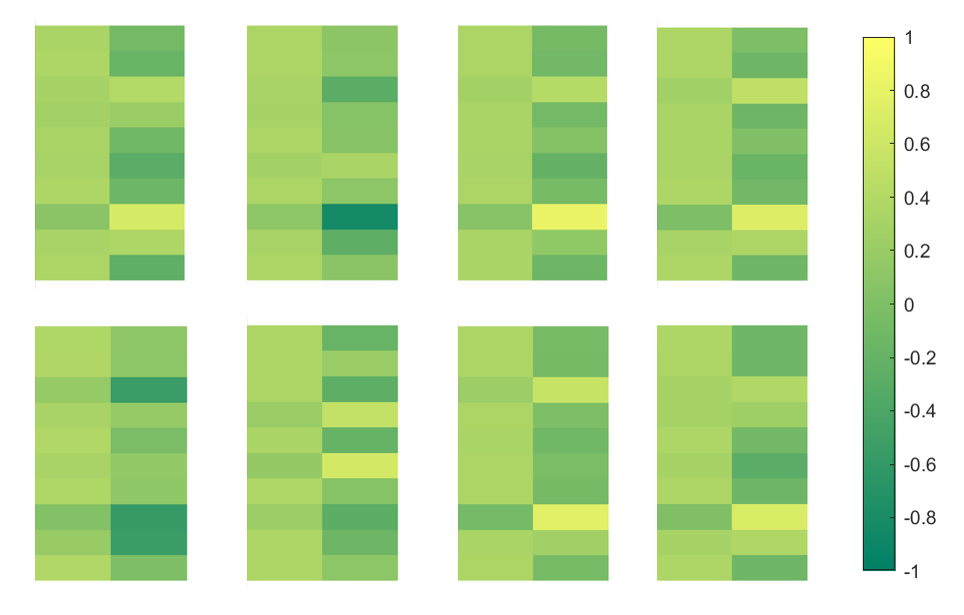}
  \caption{Principal component (PC) coefficient heatmap for NFT with two PCs. Each heatmap represents the coefficient for a single art NFT (with 2 PCs).}
  \label{222compare}
\end{figure}

In Fig. \ref{222compare}, the linear combinations for each PC in the NFTs (with 2 PCs) are comparable. From top left to top right: ArtBlocks, BoredApeYachtClub, CloneX, CoolCats. From bottom left to bottom right: Meebits, TheSandbox, VeeFriends, WorldofWoman.

\begin{figure}[!t]
\centering
  \includegraphics[scale = 0.4]{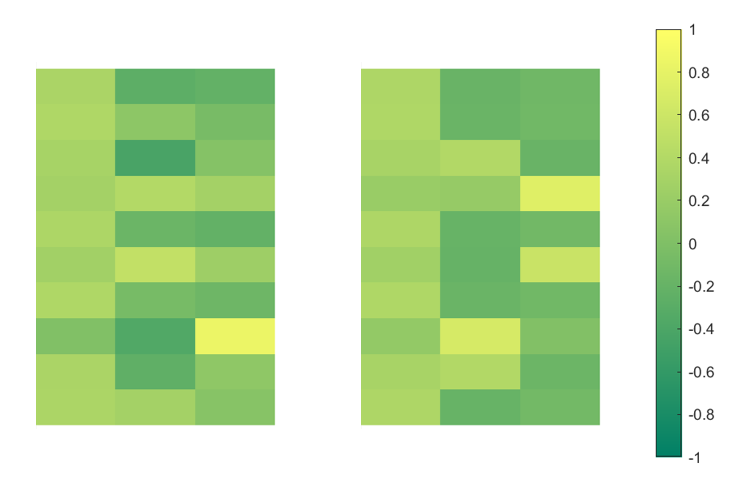}
  \caption{Principal component (PC) coefficient heatmap for NFT with three PCs. }
  \label{33compare}
\end{figure}

In Fig. \ref{33compare}, each heatmap represents the coefficient for a single art NFT (with 3 PCs). From left to right: Azuki, CrypToadz.

\subsection{Interpretation and Inferences}

CryptoPunks -- the only NFT marketplace that needs at least four PCs to reach 90\% cumulative proportion of variance explained -- shows different behaviors from the other NFT marketplaces. For our PCA interpretation, we choose the linear combination of features that significantly contributes to the model. Taking CryptoPunks as a special example, due to its relatively stable yet massive transaction amounts, we can write the following linear combination for each principal component:

\setlength{\arraycolsep}{0.0em}
\begin{eqnarray}
PC1&{} ={}  & 0.38\times SalesUSD \nonumber\\
    & {} & +0.39\times ActiveMarketWallets \nonumber\\
    & {} & +0.37\times SecondarySalesUSD  \nonumber\\
    & {} & +0.39\times UniqueBuyers \nonumber\\
    & {} & +0.39 \times UniqueSellers 
    \label{CryptoPunk1}
\end{eqnarray}

\setlength{\arraycolsep}{0.0em}
\begin{eqnarray}
PC2&{}={}&0.58\times AverageUSD
\label{CryptoPunk2}
\end{eqnarray}

\setlength{\arraycolsep}{0.0em}
\begin{eqnarray}
PC3&{} ={}  & 0.57\times PrimarySales \nonumber\\
    & {} & +0.59\times PrimarySalesUSD 
\label{CryptoPunk3}
\end{eqnarray}

\begin{eqnarray}
PC4&{} ={}  & 0.43\times NumberOfSales \nonumber\\
    & {} & +0.52\times econdarySales 
\label{CryptoPunk4}
\end{eqnarray}

For BAYC, we have the following:

\begin{eqnarray}
PC1&{} ={}  & 0.35\times NumberOfSales \nonumber\\
        & {} &+\ 0.33\times SalesUSD \nonumber\\
        & {} &+0.35\times ActiveMarketWallets \nonumber\\
        & {} &+0.34\times SecondarySales  \nonumber\\
        & {} &+0.31\times PrimarySalesUSD \nonumber\\
        & {} &+0.32\times SecondarySalesUSD \nonumber\\
        & {} &+0.34\times UniqueBuyers \nonumber\\
        & {} &+0.35\times UniqueSellers
    \label{BAYC1}
\end{eqnarray}

\begin{eqnarray}
PC2&{} ={}  & -0.83\times AverageUSD \nonumber\\
    & {} & +0.34\times PrimarySales 
\label{BAYC2}
\end{eqnarray}

Note that from Eq. \ref{CryptoPunk4} we can see that the features of SecondarySales and NumberOfSales make a significant contribution to the fourth principal component of CryptoPunks. On the other hand, it is not the case for BAYC as shown in (\ref{BAYC1}) and (\ref{BAYC2}). From this observation, we can say that CryptoPunks is a relatively more stable market, where the secondary transaction trading sizes are substantially larger than others based on Fig \ref{saleNumSecFirst}, \ref{SalesSecFirst}, \ref{buyerSecFirst} and PCA results.

\section{Concluding Remarks}
NFTs are a new digital asset in the blockchain network. Its utility features and marketplaces are still in the process of reaching a point where users can find a more healthy and safe trading experience on digital assets. Sooner than later, NFTs may be linked to some physical counterparts (as utility NFTs). This paper finds some trends and patterns from the selected NFT trading marketplaces. Based on our data collection and analysis, the number of secondary sales is much higher than that of the primary sales. In other words, the art NFTs are still trading assets, bought and sold for short-term objectives, rather than long-term investments. Among our selected NFT marketplaces, CryptoPunks showed a unique pattern: (i) much fewer transactions; (ii) higher average sales amount. CryptoPunks is the only marketplace with four principal components for the PCA, explaining 95\% of the total variation. In contrast, other marketplaces have two or three PCs with a similar explanation level of the variation. We hope our research delivers useful summaries and insights into the art NFTs as well as their unstable yet rapidly converging marketplaces.


\end{document}